\documentclass[prb,twocolumn,superscriptaddress,showpacs]{revtex4-1}
\usepackage{graphicx}
\usepackage{epsfig}
\usepackage{amssymb,amsmath,amsfonts,hyperref}
\usepackage{wasysym}
\usepackage{latexsym}

\usepackage{color}

\newcommand{\be}{\begin{eqnarray}}
\newcommand{\ee}{\end{eqnarray}}

\begin{document}
\title{Impurity spin texture at a N\'eel-Valence Bond Solid critical point in $d=2$ SU(3) quantum antiferromagnets}
\author{Argha Banerjee}
\affiliation{Tata Institute of Fundamental Research, 1, Homi Bhabha Road,
Mumbai 400005, India}
\author{Kedar Damle}
\affiliation{Tata Institute of Fundamental Research, 1, Homi Bhabha Road, Mumbai
400005, India}
\author{Fabien Alet}
\affiliation{Laboratoire de Physique Th\'eorique, Universit\'e de Toulouse, UPS, (IRSAMC), F-31062 Toulouse, France}

\begin{abstract} We study the impurity physics at a continuous quantum phase transition from an 
SU(3) symmetric N\'eel ordered state to a valence bond solid state that 
breaks lattice symmetries, using quantum Monte Carlo techniques. This continuous transition is expected to be
an example of `deconfined criticality' in an SU(3) symmetric system. 
We find that the spin-texture induced by a 
missing-spin defect at the transition takes on a finite-size scaling 
form consistent with expectations from standard scaling arguments at a 
scale-invariant quantum critical point, albeit with significant 
subleading power-law finite size corrections that we analyze in detail. Together with recently-found logarithmic violations of scaling at similar continuous transitions
in the SU(2) case, our results provide indirect evidence for the existence of operators that become marginal as $N$ is reduced to $2$ in the field
theoretical description of these deconfined critical points.
\end{abstract}

\vskip2pc

\maketitle

\section{Introduction}

Impurity effects, arising from naturally occurring or intentionally 
introduced defects, have been exploited often as probes of the low 
temperature physics of various condensed matter systems. The 
characteristic response pattern that results from the presence of 
impurities usually contains spatially resolved information
which provides clues about the nature of the low temperature phase,
the presence of long-range order,
and the role of quantum fluctuations and other competing orders that are 
suppressed in the bulk but may be realized close to the impurity. Such 
clues are crucial in the context of materials with strong electronic 
correlations whose physics is not as well understood as ordinary metals, 
and such an approach has therefore been heavily used in experiments on 
unconventional superconductors and Mott insulating 
antiferromagnets~\cite{Alloul_RMP}.

On the theoretical front, the very nature of these strongly correlated 
systems makes detailed analytical solutions difficult even for 
simplified models. Furthermore, mean-field methods often fail to capture 
the strong correlations well enough to be reliable. In cases where 
efficient numerical computations are possible, large-scale 
high-precision numerical studies can be used to access the physics of 
such systems. Other approaches that have had some degree of success 
include `large-N' approximation schemes, in which the symmetry group of 
the physical problem (for instance, SU(2) of spin-rotation invariance) 
is enlarged (for instance replacing SU(2) by SU(N)), and the resulting 
generalization solved by approximations that become exact in the 
large-$N$ limit. This approach has been particularly useful and 
influential in the study of quantum magnetism, and the generalization of 
the SU(2) spin rotational symmetry to SU(N) has played a key role in our 
understanding of the magnetism of strongly correlated Mott insulators.

For instance, the analysis of the SU(N) Heisenberg model by Read 
and Sachdev~\cite{Read_Sachdev_NPB89, Read_Sachdev_PRB90} provided the 
first insights regarding the manner in which a N\'eel-ordered SU(2) 
antiferromagnet can be destabilized by quantum fluctuations in the
antiferromagnetic order to form a Valence Bond Solid (VBS). In their analysis,
Read and Sachdev considered a SU(N) 
generalization in which A-sublattice sites of the square lattice carry 
an SU(N) representation with $m=1$ rows and $n_c$ columns, while 
$B$-sublattice sites carried a representation with $N-m \equiv N-1$ rows 
and $n_c$ columns. Assuming short-ranged N\'eel order, but making no 
assumptions about the presence of long-range N\'eel order, 
Read and Sachdev argued that the physics of such magnets can be 
described by a CP$^{N-1}$ field theory coupled to a (compact) U(1) gauge 
field in two spatial dimensions, with a dimensionless coupling $g$ that 
is a function of both $n_c$ and the microscopic details of the interactions 
between spins (for instance, the strength of additional exchange or ring-exchange terms). This CP$^{N-1}$ theory admits point-like space-time 
`hedgehog' defects which correspond to monopoles in the compact U(1) 
gauge field. These hedgehogs carry Berry phases that depend on their spatial 
locations and the value of $n_c$. In the antiferromagnetic phase 
stabilized at small $g$, these hedgehogs are energetically 
suppressed, and their Berry phases can be ignored. Conversely, the 
VBS phase corresponds to a `condensate' of these 
hedgehogs, with their position-dependent Berry phases being responsible 
for the spontaneous breaking of spatial translational symmetry characteristic
of the VBS phase.

In an insightful paper~\cite{Senthil_etal_Science04}
Senthil {\em et al.} used this picture of the two phases to
argue that the quantum phase transition between them in two dimensional SU(2)
antiferromagnets can be generically a continuous transition. This is in
conflict with expectations from the usual Landau theory
approach, which predicts that a direct transition between these
two phases should generically be first order.
In the analysis of Senthil {\em et al.}, this transition
can be generically continuous
{\em if} quadrupled hedgehog defects\cite{footnote-quadrupole}, which are known to be irrelevant
in a renormalization group sense in the large-$N$ limit, remain irrelevant in the
$N=2$ case. As a result of the irrelevence of hedgehog defects,
the `natural' description of this
transition is then in terms of `deconfined' spinon fields of a {\em non-compact}
CP$^1$ model (NCCP$^1$) coupled to a non-compact
$U(1)$ gauge field, rather than the antiferromagnetic vector order parameter field corresponding to the N\'eel ordering
on one side of the transition or the complex order parameter corresponding to the VBS order on the other. 

This prediction of a `deconfined critical point' separating
the N\'eel and VBS phases motivated several numerical studies that tried
to test its validity in two quite different ways.
One of these involved the numerical study of specific lattice discretizations
of the NCCP$^1$ field theory itself. Using such an
approach, Kuklov {\em et al.} have argued that certain lattice
discretizations of the NCCP$^1$ field theory generically admit a weakly first-order transition~\cite{Kuklov_etal_PRL08} with finite correlation length,
rather than a critical point
with diverging correlation length. However, Motrunich and Vishwanath have 
demonstrated~\cite{Motrunich_Vishwanath_arxiv08} that other lattice 
discretizations of the NCCP$^1$ field theory do admit a critical point 
whose properties are expected to provide a good description of the 
generically continuous N\'eel-VBS transition. 

The second set of studies focused instead on microscopic spin models
capable of displaying such N\'eel-VBS transitions. The most studied
of these are the `$J-Q$ models' constructed by Sandvik~\cite{Sandvik_PRL2007},
in which a direct transition from N\'eel to VBS order
is driven by the competition between four-spin plaquette interactions ($Q$ terms) that favour VBS order on the square lattice and two-spin exchange
interactions ($J$ terms) that favour two-sublattice N\'eel order on
the square lattice. While initial studies of the $J-Q$ models~\cite{Sandvik_PRL2007,Melko_Kaul_PRL2008,Jiang_etal_JStatmech2008} did not fully resolve the
nature of this transition, more recent works\cite{Lou_etal_PRB09,Sandvik_PRL2010,Kaul_2010,Banerjee_Damle_Alet_PRB2010} have convincingly demonstrated the generically continuous nature of the N\'eel-VBS transition in this class of SU(2) symmetric $S=1/2$ models on the square lattice. 

The most recent studies of the bulk N\'eel-VBS transition\cite{Sandvik_PRL2010} as well as of impurity effects at this
transition\cite{Banerjee_Damle_Alet_PRB2010} both lead to
another remarkable conclusion: although the N\'eel-VBS transition in SU(2) symmetric $S=1/2$ systems on the square lattice is indeed continuous, logarithmic violations of scaling behaviour, of the type that
can arise when the critical theory contains marginally irrelevant operators, are also present at the critical point. These logarithmic effects could be the cause of the conflicting conclusions from the earlier numerical studies of this transition in the $J-Q$ models, with some studies\cite{Sandvik_PRL2007,Melko_Kaul_PRL2008}
interpreting their results in terms of a continuous transition with conventional scaling properties,
and other studies interpreting very similar data in terms of a weakly first-order 
transition~\cite{Jiang_etal_JStatmech2008}.

It is therefore natural to ask if this reflects the fact that quadrupled 
monopole operators, which are known to be strongly irrelevant in the large-$N$ limit\cite{Senthil_etal_Science04}, actually become 
marginally irrelevant in the physical $N=2$ theory. While this question 
cannot be directly addressed by computational studies in the absence of 
fresh insights on the field theory side, it does provide a strong 
motivation to study similar N\'eel-VBS transitions on the square lattice 
in microscopic SU(N) spin models with enlarged symmetry ($N>2$).
If these transitions are continuous, it is of great interest to see whether they obey predictions from standard scaling theory without logarithmic violations. This is the question we address in this work.

In order to do this, we focus on the SU(3) $J-Q$ 
model\cite{Lou_etal_PRB09} which has antiferromagnetic exchange 
interactions $J$ between nearest neighbours and a four-spin coupling $Q$ 
between sites belonging to a plaquette. In the language of Read and 
Sachdev\cite{Read_Sachdev_NPB89}, the spins in this model carry a 
representation with $n_c=1$ columns and $m=1$ rows on the A-sublattice 
and $n_c=1$ columns and $N-m \equiv N-1$ rows on the B-sublattice, with 
$N=3$ in the case of SU(3). This model generalizes the SU(N) 
antiferromagnetic model with purely nearest neighbour exchange that was 
studied in earlier works~\cite{Kawashima,Beach_etal_PRB09}, 
who found that the ground state is N\'eel ordered for $N =3$ and $N=4$. 
From the work of Lou {\em et. al.}~\cite{Lou_etal_PRB09} and Kaul~\cite{Kaul_2010}, it is known that this SU(3) model 
has a continuous transition between this N\'eel ordered state and a 
VBS ordered singlet ground state as $Q$ is increased 
beyond $Q_c/J\approx 0.502$.

Here, we study the zero temperature impurity 
spin texture that is induced by a single vacancy in the lattice at this SU(3)
transition point, and confront our computed results with predictions from finite-size scaling 
theory\cite{Banerjee_Damle_Alet_PRB2010,Hoglund_Sandvik_PRL07,Metlitski_Sachdev_PRB07} 
that rely on the scale-invariance of the underlying quantum critical 
point. By a careful analysis of the scaling properties of the Fourier
components of the spin texture near wavevectors ${\bf k}= {\bf 0}$, and
${\bf k}= {\bf Q}$ (where ${\bf Q} \equiv (\pi,\pi)$ is the antiferromagnetic
ordering wavevector on the square lattice), we conclude that
the N\'eel-VBS transition in such SU(3) magnets on the square lattice
obeys finite-size scaling forms expected at a scale-invariant quantum
critical point, albeit with noticeable subleading finite-size corrections
to scaling. In other words, the hitherto unidentified marginal operator
in the $N=2$ critical theory is no longer marginal in the $N=3$ case. The recent work of Kaul~\cite{Kaul_2010} adresses similar issues, albeit considering {\it bulk} properties of the transition in the SU(3) and SU(4) $J-Q$ models.

The plan of the paper is as follows. We briefly introduce the SU(N) 
$J-Q$ model in Sec.~\ref{sec:model}. We then describe the projector 
quantum Monte Carlo method used to simulate the SU(N) $J-Q$ model with a 
single non-magnetic impurity in Sec.~\ref{sec:QMC}, first
describing our choice of basis in Sec.~\ref{sec:basis}, and then outlining
how existing algorithms must be modified to account for both the 
existence of the SU(N) symmetry and of a spinless impurity. We then 
present and analyze our numerical results for the impurity-induced spin 
texture at the critical point of the SU(3) $J-Q$ model in 
Sec.~\ref{sec:results}. Finally, we discuss implications of our results 
and present our conclusions in Sec.~\ref{sec:discussion}.

\section{Model} 
\label{sec:model}

We consider a system made of $N_A$ $A$-sublattice sites with `spins' that 
carry the defining representation ${\mathbf{N}}$ of SU(N) and $N_B$ $B$-sublattice sites at which 
`spins' carry the conjugate representation ${\mathbf{N}}^{*}$. Following Refs.~\onlinecite{Kawashima,Beach_etal_PRB09,Lou_etal_PRB09}, we write the Hamiltonian as: 
\begin{equation}
  H =- J\sum_{\langle i j \rangle} P_{i j} -Q\sum_{\langle i j \rangle \langle k l \rangle} P_{ij}P_{kl},
  \label{eq:H}
\end{equation}
where $P_{i j} = -\frac{1}{N}
  {\mathcal A}^{\alpha}_{\beta}(i) {\mathcal A}^{\beta}_{\alpha}(j)$,
${\mathcal A}^\alpha_\beta$, the generators of the SU(N) algebra satisfy the 
relations
$  [{\mathcal A}^{\alpha}_{\beta}(i),{\mathcal A}^{\mu}_{\nu}(j)]
  =
  \delta_{ij}
  \left(
  \delta_{\alpha\nu} {\mathcal A}^{\mu}_{\beta}(i)
  -
  \delta_{\mu\beta}  {\mathcal A}^{\alpha}_{\nu}(i)
  \right)
$. The $J$ term acts on links $\langle ij\rangle$
connecting nearest neighbour pairs of sites of the square lattice while
the $Q$ term acts on a pair of parallel nearest neighbour links 
$\langle i j \rangle \langle k l \rangle$ on the same plaquette of the square lattice.
As indicated above, we choose to represent these generators
in two different representations on the two sublattices of the square lattice.
On the $A$-sublattice, we use  the $N \times N$ matrices of the fundamental
representation ${\mathbf {N}}$ acting on states $|\alpha \rangle$
($\alpha=1,2,\dots N$) that live on each $A$-sublattice site, while on the $B$-sublattice,
the SU(N) generators are represented in terms of matrices of the complex conjugate 
${\mathbf {N}}^{*}$of the fundamental representation that act on the basis states
$|\bar{\alpha}\rangle$ on each $B$-sublattice site. 

The physics of this antiferromagnetic SU(N) model becomes clear upon 
noting that when $i$ and $j$ belong to different sublattices, $P_{ij}$ 
is actually a projector onto the SU(N) singlet ${\mathbf {1}}$ 
contained in the decomposition of ${\mathbf N} \otimes 
{\mathbf{N}}^{*}$. By analogy to the SU(2) case, the $J$ term is thus 
expected to favour long-range antiferromagnetic order in the SU(N) 
spins, while the Q term favours a `valence-bond' solid in which a 
particular pattern of pairs of spins lock into SU(N) singlets, 
breaking the symmetry of lattice translations.

As is well-known, this antiferromagnetic SU(N) model can also be 
thought of as in terms of SU(2) spin $S=(N-1)/2$ variables interacting 
via a Hamiltonian with enhanced $SU(2S+1)$ symmetry. More precisely, we 
make the correspondence $|\alpha \rangle \equiv |S^z = 
\alpha-S-1\rangle$ for $\alpha = 1,2, \dots N$ on the $A$-sublattice, 
and $(-1)^{\alpha-N}|\bar{\alpha} \rangle \equiv | S^z = 
S+1-\alpha\rangle$ for $\alpha=1,2,\dots N$ on the $B$-sublattice. In 
this language, the SU(N) singlet state between a $A$-sublattice spin 
and a $B$-sublattice spin $| \phi_{ij} \rangle = 
\frac{1}{\sqrt{N}}\sum_{\alpha=1}^{N}|\alpha \rangle_i \otimes 
|\bar{\alpha} \rangle_j$ can be rewritten as $|\phi_{ij}\rangle = 
\frac{1}{\sqrt{2S+1}}\sum_{m=-S}^{S}(-1)^{m-S}|S^z_i=m\rangle \otimes 
|S^z_j = -m\rangle$ and thus, the matrix elements of $P_{ij}$ are explicitly given by
\begin{equation}
  \langle m_i', m_j' |P_{ij}|
  m_i,m_j \rangle = 
  \frac{(-1)^{m_i+m_i'-2S}}{2S+1}  \   \delta_{m_i, -m_j}
  \  \delta_{m_i',-m_j'},
\end{equation}

\section{Method}
\label{sec:QMC}

\subsection{Choice of basis}
\label{sec:basis}

Our goal is to compute expectation values of local operators in the ground-state of Hamiltonian 
Eq.~\ref{eq:H} for a system with a given (odd or even) number of sites.
For the SU(2) invariant, nearest neighbour
$S=1/2$ antiferromagnet at $Q=0$ on a finite bipartite lattice, we 
know from the Lieb-Mattis theorem~\cite{LM} that the ground-state is a 
$S_{\mathrm{tot}}=0$ singlet when the total number of sites is even,
and is doubly degenerate corresponding to $S_{\mathrm{tot}}=1/2$ when the 
number of spins is odd. Unfortunately, it does not appear possible
to come up with a simple extension of the Lieb-Mattis argument
that would lead to such a result for Hamiltonian Eq.~\ref{eq:H} when 
$N>2$, even in the case with $Q=0$. The main obstruction is that the Perron-Frobenius 
theorem (at the heart of the original Lieb-Mattis proof) cannot be used 
for the Hamiltonian in the total $S^z$ basis since this Hamiltonian is 
known to be {\it reducible} in each $S^z$ sector~\cite{Parkinson}.

Even though we have been unable to prove the corresponding statement, we nevertheless expect that the 
ground-state of Eq.~\ref{eq:H} belongs to the SU(3) singlet sector when 
the number of sites is even. When the number of $A$ sublattice sites $N_A$
equals $N_B+1$,
where $N_B$ is the number of  $B$ sublattice sites, we similarly expect,
by analogy to the Lieb-Mattis result for $S=1/2$ SU(2) antiferromagnets,
that the ground state of this  antiferromagnetic Hamiltonian lies in the sector 
corresponding to the  smallest irrep of SU(N) contained in the decomposition of a
tensor  product of $N_{B}+1$ copies of ${\mathbf{{N}}}$ and $N_B$ copies of 
${\mathbf {{N}}}^*$. Since the smallest dimensionality irrep in this 
decomposition is ${\mathbf{N}}$, this then suggests that the ground 
state multiplet of an SU(N) antiferromagnet with one extra 
$A$-sublattice site will carry representation ${\mathbf {{N}}}$.
In the language of the equivalent $S=1$ system for $N=3$, this translates to
the statement that the ground state for a system with $N_A=N_B+1$ has total spin $S_{\rm tot} = 1$.

We have  checked both these expectations for the pure Heisenberg model in one dimension 
with DMRG for $L$ up to $64$, and with Green Function Monte Carlo for the 
model Eq.~\ref{eq:H} with $L=4$ (for some values of $J/Q$). In all cases studied, the
ground state was found to be in the expected sector. We further found 
(for the samples studied) that the lowest-energy levels
in all spin sectors of the equivalent spin $S=(N-1)/2$ Hamiltonian follow (a
 weaker version of) the Lieb-Mattis ordering~\cite{LM}: $E_0(S+1)\geq E_0(S)$.

For systems with an even number of sites, the SU(N) singlet sector is 
spanned by SU(N) valence bonds, where each $A$-sublattice spin forms a 
SU(N) singlet with a $B$-sublattice spin\cite{Beach_etal_PRB09}, and
the efficient projection algorithm of Sandvik and Evertz can be
readily generalized\cite{Lou_etal_PRB09,Beach_etal_PRB09} to project
out the ground state written in this basis and sample expectation
values of various operators. When 
the number of $A$ sublattice sites is larger than the number of $B$ 
sublattice sites by one, and the ground state is $N$-fold degenerate 
(and carries the fundamental representation ${\mathbf{N}}$), we can 
proceed in a manner entirely analogous to the approach followed in 
Ref~\onlinecite{Banerjee_Damle_JStatmech10}. There, the SU(2) singlet 
sector algorithm of Sandvik and Evertz was generalized to cases where the 
ground state of a SU(2) symmetric $S=1/2$ antiferromagnet has total spin 
$S_{\mathrm{tot}}=1/2$ due to the presence of one extra $A$-sublattice 
site.

To do this, we use a modified basis that leaves one A-sublattice site $a_f$ `free' (in 
an arbitrary state $|\alpha\rangle_{a_f}$), and allows the remaining 
$N_A-1$ A-sublattice spins to form SU(N) singlets with the $N_B=N_A-1$ 
B-sublattice spins. The full basis is thus obtained by allowing all 
choices of free site $a_f$, all choices of state $|\alpha\rangle_{a_f}$ 
for this free site, and all valence bond pairings possible between the 
other $N_A-1$ $A$-sublattice spins and the $N_B$ $B$-sublattice spins. 
Using an argument completely analogous to those given in 
Ref~\onlinecite{Banerjee_Damle_JStatmech10} for the SU(2) case, it is easy to 
see that this basis spans the ground-state sector of this SU(N) magnet 
with $N_B=N_A-1$ $B$-sublattice sites and $N_A$ $A$-sublattice sites 
because states from this basis can be used to construct {\em all} states of 
the SU(N) valence bond basis of Beach {\em et. al.}\cite{Beach_etal_PRB09} and Lou {\em et. 
al.}\cite{Lou_etal_PRB09} for the singlet sector of the auxillary system obtained by adding to
the odd-site system one extra $B$-sublattice spin that carries representation ${\mathbf N}^{*}$.

\subsection{Modified algorithm and estimators}
\label{sec:algorithm}

The SU(N) generalization\cite{Lou_etal_PRB09,Beach_etal_PRB09} of the Sandvik-Evertz
algorithm\cite{Sandvik_Evertz_PRB2010} can be modified to account for an odd number of sites
and a $N$-fold degenerate ground state in a very simple way, following Ref~\onlinecite{Banerjee_Damle_JStatmech10}.
In this approach, the ground-state is obtained by stochastically sampling the
action of $(-H)^m$ (with some large power $m \sim L^3$ for a $L \times L$
system of linear size $L$) on an
initial trial state $|\psi_{S}\rangle$ in the ground state sector with $S_{{\mathrm{tot}}} = S^z_{{\mathrm{tot}}} = S$ with $S=(N-1)/2$ (in the equivalent
spin representation).
More specifically, the `space-time loop' representation\cite{Sandvik_Evertz_PRB2010} of
$\langle \psi_{S}|(-H)^{2m}|\psi_{S}\rangle$ has, apart from the closed space-time loops that can
take on any one of $2S+1$ states\cite{Lou_etal_PRB09,Sandvik_Evertz_PRB2010}, exactly one extra `open-string' (as in Ref.~\onlinecite{Banerjee_Damle_JStatmech10})
with alternating $A$-sublattice and $B$-sublattice nodes which are fixed to be in the $S^z=+S$ and $S^z=-S$ state
respectively. Unlike the closed loops, the open string thus has only one allowed state. We use loop updates to flip between the $2S+1$ allowed
states of each closed loop, exactly as in the SU(N) algorithm of Ref.~\onlinecite{Lou_etal_PRB09}, but do not change the state of the open string. All other aspects
of the algorithm remain essentially unchanged, apart from the fact that valence-bond updates
are designed in our case to allow the `free site' $a_f$ to move around with weights derived from the trial state $|\psi_S\rangle$ 
written in our chosen basis. To obtain a convenient representation of the overlap
between $(-H)^m|\psi_S\rangle$ and $\langle \psi_S|(-H)^m$, we follow Sandvik and Evertz and keep track of space-time loops that
cross the central space-time `slice' to obtain the valence-bond loop representation of this overlap. However, as in Ref~\onlinecite{Banerjee_Damle_JStatmech10}, we also
need to keep track of the open string, since it necessarily crosses the central space-time slice, and therefore gives rise
to an open-string in the valence bond loop representation of this overlap.

Generalizing the computations of Ref.~\onlinecite{Banerjee_Damle_JStatmech10} to the SU(N) case, one can obtain
modified estimators for various physical quantities in terms of this valence-bond loop representation
of the overlap between  $(-H)^m|\psi_S\rangle$ and $\langle \psi_S|(-H)^m$. As we focus here on the impurity-induced spin texture, we are
interested mainly in the computation of the expectation value $\langle S^z({\mathbf{r}})\rangle$
in the ground state with total $z$ component of spin $S^z_{\rm tot}=S$. It is quite easy to see
that this takes on a very simple expression in terms of the probability, $P_{\rm open}({\mathbf{r}})$, that the open
string passes through the site ${\mathbf{r}}$:
\begin{displaymath}
\langle S^z({\mathbf{r}}) \rangle = (-1)^{{\mathbf{r}}} S P_{\rm open}({\mathbf{r}})
\end{displaymath}
This is because the contributions of all closed loops in the valence-bond overlap cancel out since a site covered
by a closed loop is equally like to be in any $S^z$ state.

Below, we use this to study the impurity induced spin-texture created by a missing-site defect at the critical
point of the SU(3) $J-Q$ model Eq.~\ref{eq:H}.

\section{Spin texture at critical point}
\label{sec:results}

\subsection{Definition and measurements of the spin texture}

The impurity spin-texture $\langle S^z({\mathbf{r}})\rangle$ in the ground state with $S^z_{\rm tot} = S$ has an alternating part
$\Phi^n({\mathbf r})$ that oscillates in sign between the two sublattices of the square lattice, and a uniform part
$\Phi^u({\mathbf r})$ that
varies slowly in space. It is therefore useful to consider the Fourier transform
\begin{equation}
S^z({\mathbf k}) = \sum_{{\mathbf r}}\langle S^z({\mathbf{r}})\rangle \exp(i {\mathbf k} \cdot {\mathbf r}) \; ,\nonumber
\end{equation}
defined for wavevectors ${\mathbf k} = 2 \pi
{\mathbf m}/L$  (where ${\mathbf m} \equiv (m_x, m_y)$, with $m_{x/y} =
0,1, \dots L-1$). This two-component description of the real space texture
implies that $S^z({\mathbf{k}})$ is expected to have two peaks, one at ${\mathbf k} = 0$
of magnitude $S$ (reflecting the fact that $S^z_{\rm tot} = +S$ in the ground state under
consideration), and the  other at ${\mathbf k} = {\mathbf Q} \equiv (\pi,\pi)$ reflecting
the tendency to N\'eel order.

\begin{figure*}[!]
{\includegraphics[width=2\columnwidth]{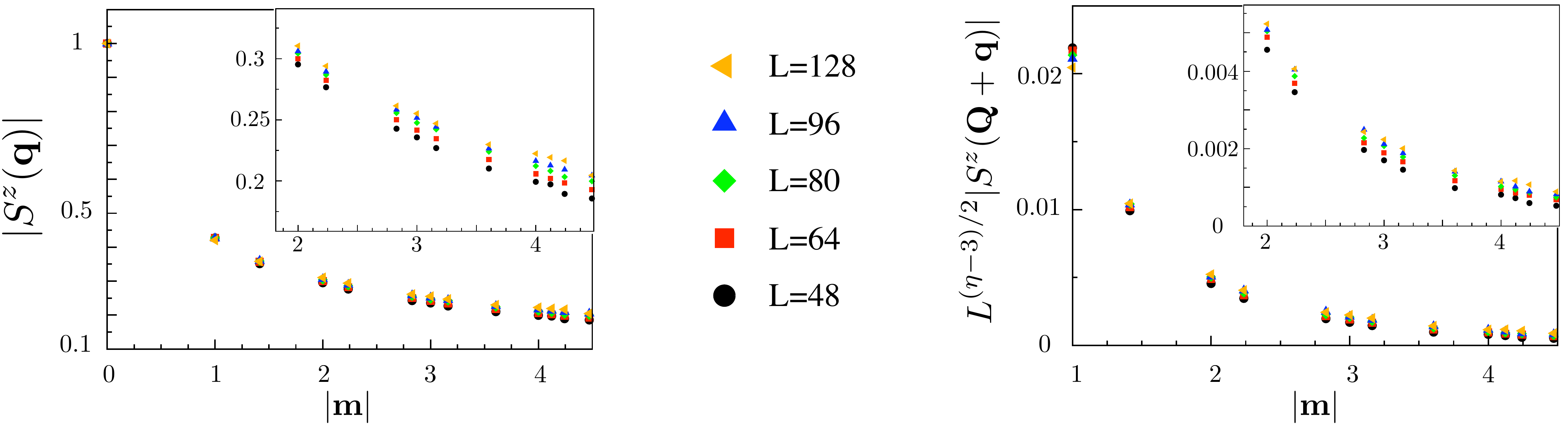}}
\caption{${\mathbf m}={\mathbf q}L/2\pi$ dependence of $|S^z({\mathbf k})|$ 
at the N\'eel-VBS transition in the SU(3) symmetric $J-Q$ model, for ${\mathbf k} = {\mathbf Q} + {\mathbf q}$ (right panel) and ${\mathbf k} = {\mathbf q}$ (left panel). The value of the bulk exponent $\eta= 0.18$ used here is obtained from a fit
of the $L$ dependence of $|S^z({\mathbf Q})|${\protect{\cite{footnote2}}}.
Insets zoom in on the small deviations from perfect scaling.}
\label{Fig1}
\end{figure*}

If the underlying quantum critical point is scale-invariant, one expects
that both the uniform and the alternating parts of the real-space spin texture obey
the predictions of standard finite-size scaling theory which assumes
that the system size $L$ is the only length-scale left in the problem at criticality.
One therefore expects\cite{Hoglund_Sandvik_PRL07,Metlitski_Sachdev_PRB07}
\be
\Phi^u({\mathbf r}) & = &\frac{1}{L^2}f^u({\mathbf r}/L) \nonumber \\
\Phi^n({\mathbf r}) &= & \frac{1}{L^{(1+\eta)/2}}f^n({\mathbf r}/L) \nonumber 
\ee
where $\eta$ is the bulk anomalous exponent associated with the N\'eel order parameter, $f^u$
and $f^n$ are the scaling forms for the uniform and alternating signals,
and this scaling form is expected to work at all $|{\mathbf{r}}| \gg 1$.
Since it is difficult to unambiguously identify the alternating and uniform
parts of the texture in real space, we find it convenient to go to Fourier space and note that these scaling expectations imply that $S^z({\mathbf k})$ obeys
corresponding scaling laws in the vicinity of these two peaks:
\be
S^z({\mathbf q}) & = & g_0({\mathbf m}) \; \; {\mathrm{for}} \; |{\mathbf q}| \ll \pi/2 \nonumber \\
S^z({\mathbf Q} + {\mathbf q}) &= &L^{(3-\eta)/2}g_{{\mathbf Q}}({\mathbf m}) \; \; {\mathrm{for}} \; |{\mathbf q}| \ll \pi/2 \; \; ,
\label{qspaceansatz}
\ee
where ${\mathbf m} \equiv {\mathbf q}L/2\pi$.

Our strategy is thus straightforward. We use the projector
quantum Monte Carlo technique to obtain high-precision data on the
impurity spin texture induced by a single vacancy in $L \times L$
samples with $L$ even and periodic boundary conditions. We use the latest estimate~\cite{Kaul_2010} of the transition point $\left. J/Q\right|_c\sim 1.9908$ (we have checked that our analysis holds for other slightly different estimates of $\left. J/Q\right|_c$).  We analyze
the corresponding Fourier transform $S^z({\mathbf k})$ near the
peaks at ${\mathbf k} = 0$ and ${\mathbf k} = {\mathbf Q}$
and ask if the scaling expectations outlined above are met.

We begin our analysis with the peak at zero wavevector, since
this part of the analysis does not even need knowledge of the
bulk anomalous exponent $\eta$. We focus on $|S^z({\mathbf q})|$
and consider its dependence on ${\mathbf m} = {\mathbf q}L/2\pi$
for various $L$ ranging from $L=48$ to $L=128$. The hypothesis of
scale-invariance requires that all these data should define
a single universal function of ${\mathbf m}$. In other words,
at fixed ${\mathbf m}$, data from various system sizes
$L$ should fall on top of each other. Our results are shown
in the left panel of Fig~\ref{Fig1} which plots  $|S^z({\mathbf q})|$ as a function
of $|{\mathbf m}|$ (the data is averaged over all available ${\mathbf m}$ that
correspond to a given  $|{\mathbf m}|$).

We see that the numerical data does seem to scale reasonably well. However,
when we zoom in (inset to Fig~\ref{Fig1}), small deviations from
perfect scaling are
visible. Next we investigate the behaviour near wavevector ${\mathbf Q}$,
using the best-fit value $\eta=0.18$ obtained by fitting $|S^z({\mathbf Q})|$\cite{footnote2}, and ask if $|S^z({\mathbf Q} + 2\pi {\mathbf m}/L)|$ collapses onto a single curve when all the data for
various sizes $L$ is plotted against ${\mathbf m}$. This is shown in the right panel of Fig~\ref{Fig1}.
Again,  we see that the data does scale reasonably well, but deviations from perfect scaling are visible upon zooming in (inset to Fig.~\ref{Fig1}). For both these peaks,
we also note that the deviations from scaling are noticeably smaller
than in the corresponding peaks for the SU(2) case (see Fig. 1 of Ref.~\onlinecite{Banerjee_Damle_Alet_PRB2010}). However, a detailed analysis
of these small deviations from scaling is needed to explore this
in more quantitative terms.

\subsection{Finite-size effects in the spin texture}

\begin{figure*}[!]
{\includegraphics[width=2\columnwidth]{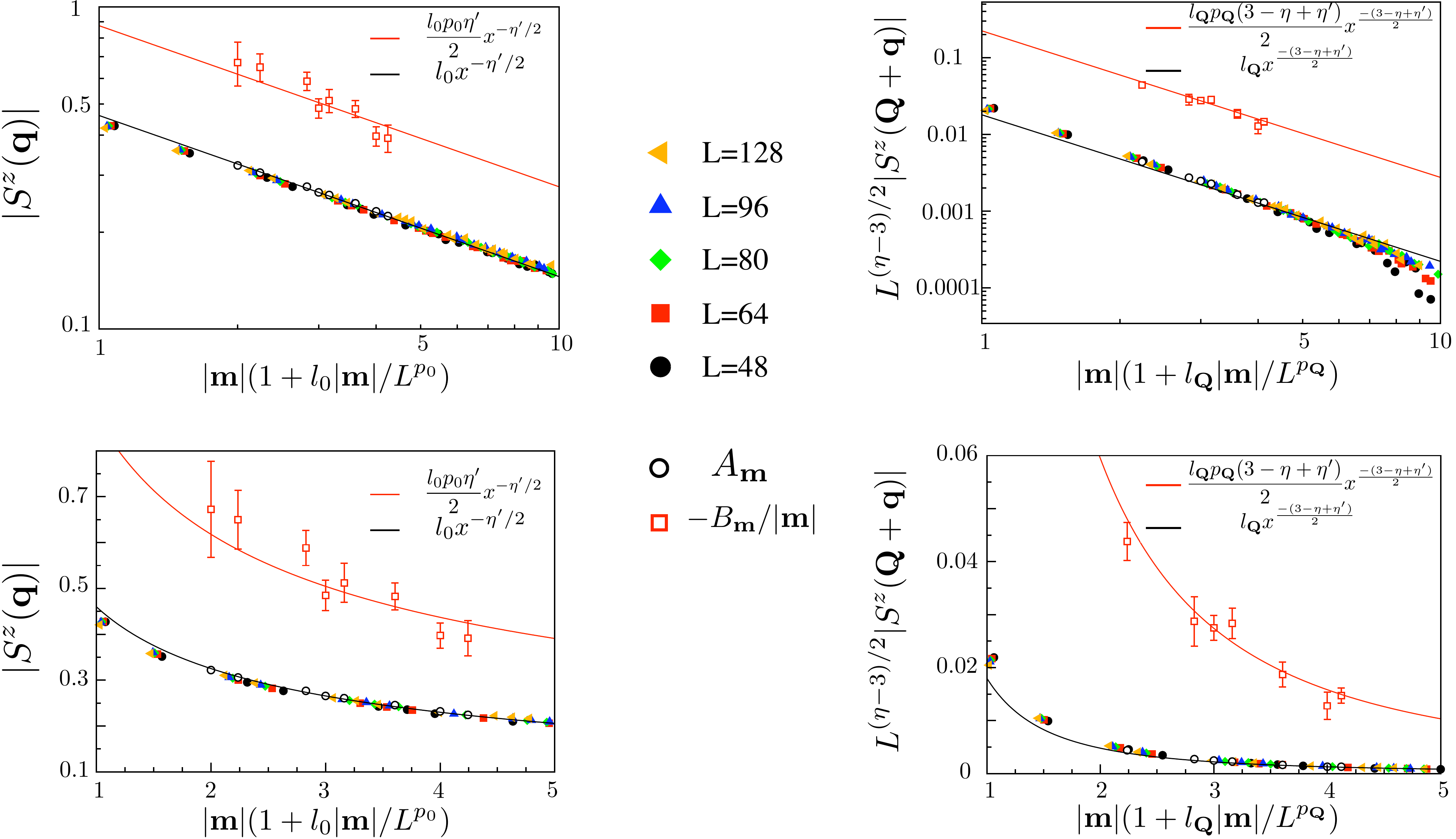}}
\caption{Left panel: Scaling collapse of $|S^z({\mathbf k})|$ 
as a function of ${\mathbf m} =  {\mathbf q}L/2\pi$ at the N\'eel-VBS transition in the SU(3) symmetric $J-Q$ model for various $L$ in accordance
with the ansatz Eqn~{\protect{\ref{qspacemodified}}}
for ${\mathbf k} = {\mathbf Q}+{\mathbf q}$ (right panels)
and ${\mathbf k} = {\mathbf q}$ (left panels). Also shown are points
representing $A_{{\mathbf m}}$, which fall right on top of the scaling
functions $g_0$ and $g_{{\mathbf Q}}$ defined by the scaling collapse. We also display the $|{\mathbf m}|$ dependence of $B_{\mathbf m}$ and compare it against
the expected dependence required by self-consistency. Data in
the vicinity of both peaks are shown in linear scale (bottom panels) as well as logarithmic
scale (top panels). Straight lines on the log-log plots display the power-law nature of the scaling
functions, with both scaling functions controlled by the single impurity scaling
exponent
$\eta^{\prime}$ as indicated in the text. The linear scale of the bottom panels highlights the quality of collapse.}
\label{Figure2}
\end{figure*}

It is important to understand these deviations from scaling,
and ask if they can be understood in terms of subleading finite-size corrections to scaling
of the standard form expected at a scale invariant quantum critical point. In order to do
this, we note that scaling theory would predict that the dimensionless argument
${\mathbf m}$ should acquire sub-leading corrections that transform it to $(1+\delta_0({\mathbf m})(l_0/L)^{p_0}){\mathbf m}$,
where the power $p_0>0$ is a property of the renormalization group flows towards the fixed
point describing this scale-invariant critical point, $\delta_0({\mathbf m})$ is a function of the dimensionless variable ${\mathbf m}$, and $l_0$ is a microscopic length scale
that depends on the details of the Hamiltonian under study. A similar
ansatz would be expected to hold for the peak centered at the antiferromagnetic
wavevector, with $\delta_0$ replaced by $\delta_{{\mathbf Q}}$, $l_0$ replaced by $l_{{\mathbf Q}}$ and $p_0$ replaced by $p_{{\mathbf Q}}$. In other words,
we modify Eq.~\ref{qspaceansatz} to now read:
\be
S^z({\mathbf q}) & = & g_0({\mathbf m}(1+l_0^{p_0}\delta_0({\mathbf m})/L^{p_0}))\nonumber \\
S^z({\mathbf Q} + {\mathbf q}) &= &L^{\frac{3-\eta}{2}}g_{{\mathbf Q}}({\mathbf m}(1+l_{{\mathbf Q}}^{p_{{\mathbf Q}}}\delta_{{\mathbf Q}}({\mathbf m})/L^{p_{{\mathbf Q}}})) \; ,
\label{qspacemodified}
\ee
for $|{\mathbf q}| \ll \pi/2$. 
Any such ansatz that incorporates subleading finite-size effects
must satisfy three conditions:
First, the data for various sizes, when plotted against the scaling
argument suggested by such an ansatz, must appear to collapse onto a
single curve that defines the corresponding scaling function $g$.
Second, the $L$ dependence of $S^z$ at each value of ${\mathbf m}$
should fit well to the form $A_{{\mathbf m}}+B_{{\mathbf m}}/L^p$ in which
$p$ is fixed by the choice made for the scaling variable earlier.
Third, $A_{{\mathbf m}}$ and $B_{{\mathbf m}}$ must be consistent with expectations derived from the choice of
scaling variable. More explicitly,
$A_{\mathbf m} = g({\mathbf m})$, and $B_{{\mathbf m}} = l \delta({\mathbf m}){\mathbf m}\cdot {\mathbf \nabla}_mg$. These three conditions hold at
both peaks, and the subscripts indicating the peak in question can be temporarily omitted from $g$, $\delta$, $l$ and $p$.

Before we use this framework to analyze our data, two points
need to be made. The first is that the range of sizes available
is not large enough to allow a unique determination of the best
fit values for $A_{{\mathbf m}}$, $B_{{\mathbf m}}$ and $p$ from fits
to the $L$ dependence at each
${\mathbf m}$.  This is because the deviations from perfect scaling
are rather small, and as a result, one needs a very large range
in $L$ to pin down the best fit $p$. With our limited range
in $L$, a range of $p$ gives equivalently good fits for the $L$
dependence at each ${\mathbf m}$. Each choice of $p$ in this range fixes a best-fit
value of $A_{{\mathbf m}}$ and $B_{{\mathbf m}}$ when one fits
the $L$ dependence at individual ${\mathbf m}$.

\begin{figure*}[!]
{\includegraphics[width=2\columnwidth]{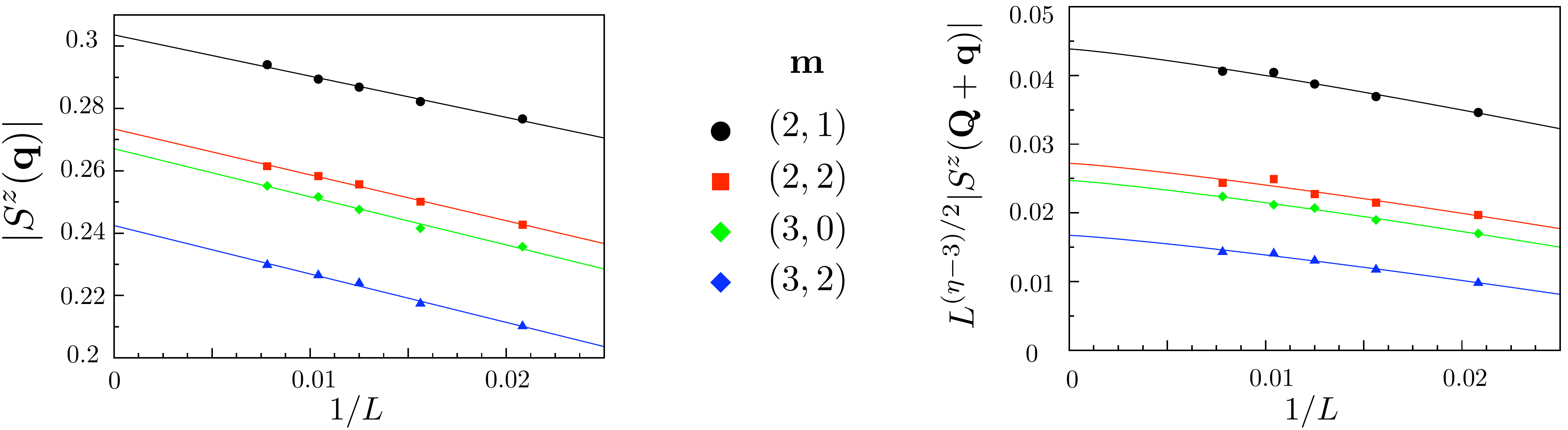}}
\caption{$L$ dependence of $|S^z(2\pi{\mathbf m}/L)|$ (left panel) and of $|S^z({\mathbf Q}+2\pi{\mathbf m}/L)|$ (right panel) 
at the N\'eel-VBS transition in the SU(3) symmetric $J-Q$ model at various ${\mathbf m}$. Solid lines
denote fits to the form mentioned in text, and the legends indicate
${\mathbf m} = (m_x,m_y)$ corresponding to each curve.}
\label{Figure3}
\end{figure*}
Likewise, the deviations from perfect scaling are
too small to uniquely fix $p$, $\delta$ and $l$ by demanding
the best scaling collapse for all the data plotted against
the scaling variable corresponding to these choices for $p$, $\delta$
and $l$. One obtains scaling collapse that looks equally good
to the eye for a range of $p$, and for at least two simple choices for
the function $\delta$: $\delta({\mathbf m}) = 1$ and $\delta({\mathbf m}) = |{\mathbf m}|$. The best one can do is to fix $l$ uniquely
after choosing a $p$ within this range and making one
of these two choices for $\delta$.

In view of this uncertainty, the self-consistency conditions outlined above
(which relate the best-fit $A_{{\mathbf m}}$ and $B_{{\mathbf m}}$
to the choices made for $\delta$ and $l$ which define the corresponding scaling
variable, and to the function $g$ defined by the resulting scaling collapse) are particularly useful in fixing the form of the
finite-size corrections from the available data. We have therefore
systematically scanned over the range of $p$ that yields a viable
description for both simple
choices of the function $\delta$. For each choice of $\delta$ and $p$,
we have determined the value of $l$ that yields the best scaling collapse,
as well as obtained the best-fit $A_{{\mathbf m}}$ and $B_{{\mathbf m}}$ 
from fits to the $L$ dependence at each individual ${{\mathbf m}}$.
We have then tested for self-consistency as outlined above.

Based on this extensive study, we conclude that
finite size corrections to perfect scaling take
on the form corresponding to $\delta_0({\mathbf m}) = \delta_{{\mathbf Q}}({\mathbf m}) = |{\mathbf m}|$, $l_0=3.8(1)$, $l_{{\mathbf Q}}=6.5(1)$, $p_0 =1.0(2)$, and
$p_{{\mathbf Q}} = 1.2(1)$. Furthermore, the scaling functions $g_0(x)$ and $g_{{\mathbf Q}}(x)$
both take on power-law forms for not too small $x$, consistent
with the expectations from the scaling theory for impurity response~\cite{Hoglund_Sandvik_PRL07}
which fixes the power-law forms of both functions in terms of
a single impurity exponent $\eta^{\prime}$:
$g_0({\mathbf m}) \sim |{\mathbf m}|^{-\eta^{\prime}/2}$ and $g_{{\mathbf Q}}({\mathbf m}) \sim |{\mathbf m}|^{-(3-\eta+\eta^{\prime})/2}$.

This scaling collapse is shown in Fig.~\ref{Figure2},
while the corresponding fits  to the data at fixed ${\mathbf m}$,
from which the values of $A_{{\mathbf m}}$ and $B_{{\mathbf m}}$
are obtained, are
shown in Fig.~\ref{Figure3}. The $A_{{\mathbf m}}$ thus
obtained are marked in Fig.~\ref{Figure2} and
seen to fall right on top of the composite curves produced
by the data collapse which define the corresponding scaling function $g_0$
and $g_{{\mathbf Q}}$. These scaling functions are fit to
power law forms expected from the scaling theory for the
impurity spin texture, and both scaling functions are seen to
be well-described by power-laws determined by a {\em single} impurity exponent $\eta^{\prime} = 1.0(1)$, as expected from the scaling analysis.
Also shown in Fig.~\ref{Figure2} is the comparison between
the best fit $B_{{\mathbf m}}$ obtained from the fits in Fig.~\ref{Figure3}
and their expected $|{\bf m}|$ dependence obtained from the self-consistency. Again,
we see that the $|{\bf m}|$ dependence of $B_{{\mathbf m}}$ is consistent
with that expected from the consistency condition.

Thus, conventional finite-size corrections
to scaling appear to give a completely consistent account of our data
for both peaks in the Fourier transform of the impurity spin texture.
In this context, it is interesting to note that
multiplicative corrections (which amount to violations of scaling)
appear to be favoured over such conventional additive corrections
to scaling by the
author of Ref.~\onlinecite{Kaul_2010} when describing the behaviour
of the spin stiffness at the transition in the pure system.

\section{Discussion}
\label{sec:discussion}
The fact that the $L$-dependence of the impurity-induced spin texture of the SU(3) model studied there can be described well in terms of ordinary finite-size corrections
to scaling should be contrasted with the findings of our recent study~\cite{Banerjee_Damle_Alet_PRB2010}
of the spin texture in the SU(2) symmetric $S=1/2$ $J-Q$ models on
the square lattice. There, we found that any attempt to fit the deviations from
scaling to this conventional form resulted in unphysically large $l$ and very
small $p$ in the SU(2) case. As a result, one was forced in the SU(2) case to consider either an unconventional form of scaling argument proportional
to ${\mathbf m}/L^p$ with very small $p$, or admit the possibility of logarithmic violations
of scaling, as  reflected in an argument of the form ${\mathbf m}/\log(L/l)$ for the scaling
function. These two possibilities are
in fact numerically indistinguishable from each other in the SU(2) case over the range
of available sizes. Since
the former has no known basis while the latter is a known to be a possible consequence of marginally irrelevant operators
in the renormalization group description of phase transitions, the most natural conclusion
consistent with the numerical evidence was that there are logarithmic violations of scaling
due to the presence of a marginal operator in the SU(2) case~\cite{Banerjee_Damle_Alet_PRB2010, Sandvik_PRL2010}. 
That this is not the case at the SU(3) critical point is the central
message of our work.

Our results thus support the scenario that some operator that is irrelevant at fixed points describing
N\'eel-VBS transitions for $N\geq 3$ becomes marginal at or near $N=2$. We hope these results will motivate further study of the field-theoretical description of the critical theories
underlying these interesting, but still only partially understood, quantum phase transitions.

\section{Acknowledgements}

We acknowledge computational resources of TIFR, 
GENCI-CCRT (Grant x2010-100225) and Calmip, as well as funding
from the Indian DST grant DST-SR/S2/RJN-25/2006 (KD) and French ANR program ANR-08-JCJC-0056-01 (FA). Our implementation of the QMC algorithm described in Section~\ref{sec:algorithm} used the ALPS libraries~\cite{ALPS}.


\begin{thebibliography}{999}
\bibitem{footnote-quadrupole} Only quadrupled hedgehog defects and their bound states
can be important at long length scales as other hedgehog defects carry an oscillating
Berry phase which renders them irrelevant for the long-distance physics.
\bibitem{footnote2}  This value of $\eta$ lies slighlty outside
the error bars of the higher value quoted in Ref.~{\protect{\onlinecite{Lou_etal_PRB09}}}. This is possibly due to the different estimate of the critical point $\left. J/Q\right|_c$, since we find that the original estimate of
the critical point in Ref.~{\protect{\onlinecite{Lou_etal_PRB09}}} yields
a higher value consistent with the original estimate of $\eta$.
\end{thebibliography}
\end{document}